\documentclass[twocolumn,aps,prd,superscriptaddress,nofootinbib,floatfix]{revtex4-2}

\usepackage{amsmath,amssymb,bm}
\usepackage{booktabs}
\usepackage{array}
\usepackage{graphicx}
\usepackage{xcolor}
\usepackage[hidelinks]{hyperref}
\usepackage{url}

\newcommand{\mpl}{M_{\rm Pl}}

\newcommand{\dd}{\mathrm{d}}
\newcommand{\yr}{\mathrm{yr}}

\newcommand{\chisq}{\chi^2}
\newcommand{\beq}{\begin{equation}}
\newcommand{\eeq}{\end{equation}}
\newcommand{\doilink}[1]{\href{https://doi.org/#1}{doi:#1}}
\newcommand{\arxivlink}[1]{\href{https://arxiv.org/abs/#1}{arXiv:#1}}

\begin{document}

\title{Clock--WEP Tests of String-Derived Dilaton and Modulus Sectors}

\author{Ahmed Farag Ali}
\thanks{Corresponding author: \href{mailto:aali29@essex.edu}{aali29@essex.edu}}
\affiliation{Essex County College, 303 University Ave, Newark, NJ 07102, USA}
\affiliation{Department of Physics, Faculty of Science, Benha University, Benha 13518, Egypt}

\date{\today}

\begin{abstract}
String compactifications and runaway-dilaton models can contain light scalar directions. Motion along these directions changes gauge couplings, the QCD scale, quark masses, and Yukawa parameters. A given string-derived sector then predicts low-energy slopes. Its gauge kinetic functions, matter-metric or Yukawa slopes, source charge, scalar trajectory, and scalar range determine a Damour--Donoghue WEP vector and a clock-drift vector. The analysis uses this map to compare such sectors with weak-equivalence-principle (WEP) and clock data. The numerical fit assumes one effective long-range scalar, or an aligned direction in a multi-field space. With that assumption, WEP tests and secular clock drifts probe related projections of the same scalar-slope vector. The minimal two-charge Damour--Donoghue fit uses Ti--Pt MICROSCOPE, Be--Ti torsion-balance, and $^{85}$Rb--$^{87}$Rb atom-interferometer data, then adds the Yb$^+$ E3/E2 secular $\alpha$-drift row. The WEP-only fit gives $D_{\hat m}=(-6.06\pm34.91)\times10^{-12}$ and $D_e=(8.80\pm46.46)\times10^{-12}$, with correlation $-0.99958$. The clock input is $\dot\alpha/\alpha=1.0(1.1)\times10^{-18}\,{\rm yr}^{-1}$. It matches the WEP marginal reach when $|\zeta|=2.37\times10^{-8}\,{\rm yr}^{-1}$, with $\zeta=\dot\varphi/\alpha_E$. For the benchmark $\dot\varphi=\xi H_0$ and $\alpha_E=10^{-3}\alpha_0$, this gives $\xi/\alpha_0\simeq0.34$. This value marks equal information between the clock row and the WEP marginal constraint. A model becomes testable after it supplies $r=D_{\hat m}/D_e$, the scalar amplitude, and $\zeta$. As an example, the Damour--Piazza--Veneziano runaway-dilaton estimate lies near $\eta\sim10^{-12}$ for order-unity $n=2$ parameters, well above the MICROSCOPE-level WEP scale. In the representative-isotope two-charge model, a MICROSCOPE-level Be--Ti-like row would reduce the long WEP axis by a factor $27.8$; a repeated Ti--Pt-like row would barely change it. The calculation tests compactification-level predictions for string-derived dilaton or modulus sectors after the model supplies its low-energy data.
\end{abstract}

\maketitle

\section{Motivation and main result}

String compactifications and runaway-dilaton models often contain scalar directions that remain light at low energy. Once a compactification is fixed, these directions are constrained by gauge kinetic functions, matter metrics, Yukawa dependences, moduli stabilization, and the scalar potential \cite{Vafa2005,OoguriVafa2007,Palti2019,DamourRunaway2002,KaplanWise2000,GKP2002,KKLT2003}. Their motion can shift $\alpha$, $\Lambda_{\rm QCD}$, quark masses, the electron mass, or Yukawa couplings. Clocks then search for drifts of constants. Free-fall experiments search for composition dependence.

The calculation below tests the low-energy scalar sector of a specified string-derived construction. The model must give the gauge kinetic functions, matter slopes, source charge, scalar path, and scalar range. These inputs place the model in a WEP--clock likelihood. Current data then bound that point or curve. The numerical part of this paper uses WEP data and one secular fine-structure drift row. Molecular spectroscopy and clock networks enter by the same rules once their time templates are specified.

This work provides an effective-field-theory likelihood and a model-to-observable map; it is not an explicit compactification calculation or a global exclusion of string-derived sectors. Every numerical reach quoted below is conditional on the stated long-range, one-direction, source-charge, and charge-truncation assumptions.

The calculation uses the two leading Damour--Donoghue composition coefficients, $(D_{\hat m},D_e)$, and adds the clock row through
\beq
 A_\alpha\equiv\frac{\dot\alpha}{\alpha}=\zeta D_e,
 \qquad \zeta\equiv\frac{\dot\varphi}{\alpha_E}.
\eeq
The relation has a limited domain. It assumes one effective rolling scalar, or an aligned multi-field trajectory. It also assumes a scalar range long enough for the WEP source-test-body scale, or a model-supplied range-dependent source charge. With these assumptions, WEP data measure source-weighted scalar couplings, while clocks measure the same microscopic slopes times the rolling rate.

Using Ti--Pt MICROSCOPE, Be--Ti torsion-balance, and $^{85}$Rb--$^{87}$Rb atom-interferometer data, the WEP-only fit gives
\begin{align}
D_{\hat m}&=(-6.06\pm34.91)\times10^{-12},\nonumber\\
D_e&=(8.80\pm46.46)\times10^{-12},
\label{eq:wepfitintro}
\end{align}
with
\beq
\Sigma_{\rm WEP}=10^{-24}
\begin{pmatrix}
1218.37&-1620.99\\
-1620.99&2158.47
\end{pmatrix},
\label{eq:wepcovintro}
\eeq
$\rho=-0.99958$, and $\chi^2_{\min}=0.177$ for one degree of freedom. The central values are consistent with zero. The useful output is the covariance: one short axis and one very long axis.

The clock row uses the present secular fine-structure drift bound from the Yb$^+$ E3/E2 comparison,
\beq
 \frac{\dot\alpha}{\alpha}=1.0(1.1)\times10^{-18}\,\yr^{-1},
\label{eq:langeresultintro}
\eeq
instead of the older Al$^+$/Hg$^+$ reference value. The equal-information threshold is
\beq
|\zeta|\gtrsim \zeta_{\rm th}
=\frac{\sigma_{\dot\alpha/\alpha}}{\sigma_{D_e}^{\rm WEP}}
=2.37\times10^{-8}\,\yr^{-1}.
\label{eq:thresholdintro}
\eeq
The threshold means that the clock row has the same marginal reach in $D_e$ as the WEP covariance. Exclusion of $\zeta$, $\xi$, or a model amplitude requires a predicted scalar amplitude.

For the benchmark conversion
\beq
\dot\varphi=\xi H_0,
\qquad
\alpha_E=10^{-3}\alpha_0,
\label{eq:rollintro}
\eeq
with $H_0\simeq7.0\times10^{-11}\,\yr^{-1}$,
\beq
\zeta=7.0\times10^{-8}\left(\frac{\xi}{\alpha_0}\right)\yr^{-1}.
\label{eq:zetaintro}
\eeq
Thus the present secular $\alpha$-clock row reaches the WEP marginal scale at $\xi/\alpha_0\simeq0.34$ in this benchmark. A tenfold clock improvement moves the value to $\xi/\alpha_0\simeq0.034$. The choice $\alpha_E=10^{-3}\alpha_0$ is a reference source charge. A complete scalar model must compute $\alpha_E$ from Earth-like matter and from the microscopic couplings that also set the WEP coefficients.

The covariance also tells us which material contrast matters. In the representative-isotope two-charge model, one MICROSCOPE-level Be--Ti-like row would reduce the long WEP axis from $58.10\times10^{-12}$ to $2.09\times10^{-12}$, a factor $27.8$. A second Ti--Pt-like row leaves it at $58.09\times10^{-12}$. Direction in charge space matters as much as acceleration precision.

Here and below, ``equal information'' means equality between the conditional clock uncertainty $\sigma_\alpha/|\zeta|$ and the marginal WEP-only uncertainty $\sigma_{D_e}^{\rm WEP}$ in the adopted two-parameter likelihood. It is a diagnostic of experimental complementarity along the selected scalar direction; it is neither a model-independent exclusion criterion nor an equivalence of the clock and WEP constraints throughout a general multi-field model space.

\section{Effective theory and scope}

Let $\varphi^a$ denote dimensionless scalar coordinates in Planck units. A low-energy action is
\begin{align}
S=&\int \dd^4x\sqrt{-g}\bigg[\frac{\mpl^2}{2}R
-\frac{\mpl^2}{2}G_{ab}(\varphi)\partial_\mu\varphi^a\partial^\mu\varphi^b
-V(\varphi)\nonumber\\
&\qquad +{\cal L}_{\rm SM}\big(g_A(\varphi),y_i(\varphi),v(\varphi);\psi,A_\mu,H,g_{\mu\nu}\big)\bigg].
\label{eq:action}
\end{align}
The logarithmic slopes are
\beq
 C_X{}^a\equiv \frac{\partial\ln X}{\partial\varphi^a},
 \qquad X\in\{\alpha,\Lambda_{\rm QCD},m_e,\hat m,\delta m,\ldots\}.
\label{eq:slopes}
\eeq
For a rolling background,
\beq
\frac{\dd\ln X}{\dd t}=C_X{}^a\dot\varphi_a.
\label{eq:driftgeneric}
\eeq
This relation is the common source of clock drift and time-dependent Standard-Model parameters. A chosen stabilization scheme, runaway model, or ultraviolet completion fixes the slopes. Effective field theory gives the low-energy dictionary. The ultraviolet boundary data enter from the model \cite{Burgess2020,Uzan2011,Uzan2025}.

We use the following notation throughout. The symbol $\alpha$ denotes the fine-structure constant. The symbols $\alpha_A$ and $\alpha_E$ denote scalar charges of a test body and the source body. The symbol $\alpha_0$ is a reference scalar-strength scale used in the benchmark conversion $\alpha_E=10^{-3}\alpha_0$.

The WEP rows also assume long-range scalar exchange. For a finite scalar mass the interaction carries a Yukawa factor,
\beq
 V_\varphi(r)\propto \frac{\alpha_A\alpha_E}{r}\,e^{-m_\varphi r}.
\eeq
The simple Earth-source coefficient $\alpha_E$ applies when this factor is essentially unity on the experimental length scale. A finite-range, screened, or environmental scalar requires a range-dependent source integral.

A second restriction concerns direction in field space. In a multi-field scalar space, the clock sector measures
\beq
 A_\alpha\equiv\frac{\dot\alpha}{\alpha}=C_e{}^a\dot\varphi_a,
\label{eq:multiclock}
\eeq
whereas WEP exchange measures source-contracted products of the form
\beq
 D_e \sim \alpha_E{}^a C_{e,a}.
\label{eq:multiwep}
\eeq
Equations \eqref{eq:multiclock} and \eqref{eq:multiwep} reduce to $A_\alpha=\zeta D_e$ for one light scalar, or for an aligned multi-field trajectory satisfying $\dot\varphi^a=\zeta\alpha_E{}^a$ in the relevant field-space metric.

This alignment condition is an explicit model restriction. Generic compactifications can have a nontrivial scalar kinetic metric, stabilization-induced mass mixing, and source-dependent screening; these ingredients may rotate the rolling, clock-sensitive, and WEP-sensitive directions relative to one another. The scalar relation used in the numerical module therefore applies only after the candidate model has shown that one light direction dominates, or has provided the aligned field-space data.

The effect of misalignment is easy to see in a two-vector picture. If the rolling direction and the source-charge direction form an angle $\theta$, the one-direction relation is rescaled by the projection between them. A $30^\circ$ angle changes the inferred relation by about $13\%$; a nearly orthogonal trajectory suppresses the link. The numerical results below are one-direction bounds.

The Higgs direction and the modulus direction are separate. For a fermion,
\beq
 m_f=\frac{y_f(\varphi)v(\varphi)}{\sqrt2},
\eeq
so
\beq
C_f{}^a=\partial_a\ln y_f+\partial_a\ln v.
\label{eq:fermionchain}
\eeq
For weak vector bosons,
\begin{align}
C_W{}^a&=\partial_a\ln g+\partial_a\ln v,\nonumber\\
C_Z{}^a&=\frac{g^2\partial_a\ln g+g'^2\partial_a\ln g'}{g^2+g'^2}+\partial_a\ln v.
\label{eq:gaugebosons}
\end{align}
Thus Higgs measurements probe derivatives with respect to electroweak vacuum structure. Clocks and WEP experiments probe derivatives with respect to the rolling scalar background. Collider measurements of Higgs couplings \cite{ATLAS2022,CMS2022} enter after a model specifies $v(\varphi)$ and $y_i(\varphi)$.

\section{Unified multi-observable likelihood}

The general data vector can be written as
\beq
 y_I=M_{IA}(\zeta,\omega,\Phi,\ldots)\,\Theta_A+n_I,
\label{eq:generallikelihoodrow}
\eeq
where $I$ labels an experimental row, $\Theta_A$ labels scalar-slope combinations, and $n_I$ is experimental noise. The Gaussian statistic is
\beq
 \chi^2(\Theta)=
 \sum_I\frac{\left[M_{IA}\Theta_A-y_I\right]^2}{\sigma_I^2}.
\label{eq:generalchi}
\eeq
Each observable adds one row. A WEP row has
\beq
 y_I=\eta_{AB},\qquad
 M_I=(\Delta Q_{\hat m}^{AB},\Delta Q_e^{AB},\ldots).
\eeq
A secular clock-ratio row has
\beq
 y_j=\frac{\dd}{\dd t}\ln R_j,
 \qquad
 M_j=\zeta(\Delta K_{\hat m}^{j},\Delta K_e^{j},\ldots),
\label{eq:clockrowgeneral}
\eeq
where the $\Delta K_X^j$ are sensitivity coefficients to the relevant constants. In the minimal $\alpha$ clock row used below, $\Delta K_e^j=1$ after converting the measured ratio to $\dot\alpha/\alpha$. Equation~\eqref{eq:clockrowgeneral} then reduces to $(0,\zeta)$ in the $(D_{\hat m},D_e)$ plane. This Yb$^+$ extraction assumes the adopted clock-sensitivity model. Scalar sectors that also vary $\mu$, $m_q/\Lambda_{\rm QCD}$, nuclear parameters, or Yukawa couplings require the full clock-sensitivity matrix.

Molecular spectroscopy fits the same structure. For a molecular transition,
\beq
\delta\ln\nu_m=
K_\alpha^m\delta\ln\alpha
+K_\mu^m\delta\ln\mu
+K_q^m\delta\ln(m_q/\Lambda_{\rm QCD})+\cdots .
\eeq
A secular molecular row needs a measured drift, a repeated comparison, or a time template. A high-precision value of the proton-electron mass ratio, such as the HD$^+$ result in Ref.~\cite{Patra2020}, helps fix the molecular sensitivity scale. It becomes a secular row of Eq.~\eqref{eq:generalchi} when a drift measurement, reference epoch, and sensitivity coefficients are supplied. This keeps absolute-constant measurements separate from drift constraints.

Clock-network searches for oscillatory ultralight fields, annual gravitational-potential modulation, or correlated environmental signals use the same likelihood with a different time template. Schematically,
\beq
 \delta\ln R_j(t)=
 \sum_X \Delta K_X^j d_X\,\varphi(t),
\eeq
and the assumed spectrum of $\varphi(t)$ determines the row weights. Clock networks, ultralight-dark-matter searches, and molecular measurements can be added to the same Fisher matrix \cite{Barontini2022,Sherrill2023,Filzinger2023,Patra2020}. This paper runs the long-range WEP plus secular $\alpha$-drift module. Equations~\eqref{eq:generallikelihoodrow}--\eqref{eq:generalchi} give the extension.

For a real model the slope vector may be lower-dimensional than the experimental parameter vector. If a model predicts a fixed ratio
\beq
 D_{\hat m}=rD_e,
\label{eq:rdefgeneral}
\eeq
then the two-charge likelihood becomes a one-amplitude test. If a model predicts an oscillatory scalar, the secular clock row must be replaced by a sinusoidal or stochastic template. If a model has several rolling moduli, the scalar $\zeta$ must be replaced by the field-space vectors $\dot\varphi^a$, $\alpha_E{}^a$, and $C_X{}^a$. A clock bound becomes a theory test after the ultraviolet model supplies the field-space trajectory and slopes.

\section{String-model interface: from compactification data to observables}
\label{sec:stringinterface}

A string-derived scalar sector becomes testable after compactification data are translated into low-energy slopes. For a modulus or dilaton coordinate $\Phi^a$, the gauge sector contains
\beq
 {\cal L}\supset -\frac{1}{4}\, {\rm Re}\, f_A(\Phi) F^A_{\mu\nu}F_A^{\mu\nu},
\eeq
so that
\beq
 \frac{1}{g_A^2}={\rm Re}\, f_A(\Phi),
 \qquad
 C_A{}^a\equiv \partial_a\ln g_A^2
 =-\partial_a\ln {\rm Re}\, f_A .
\label{eq:gaugefunctionmap}
\eeq
For QCD, ignoring threshold corrections for the purpose of the dictionary,
\beq
 \Lambda_{\rm QCD}=M\exp\left[-\frac{8\pi^2}{b_3 g_3^2(M)}\right],
\eeq
which gives
\beq
 C_\Lambda{}^a
 =\partial_a\ln M
 -\frac{8\pi^2}{b_3}\partial_a\left(\frac{1}{g_3^2}\right).
\label{eq:lambdamap}
\eeq
The electron and quark-mass slopes follow from the matter metrics, Yukawa functions, and electroweak vacuum dependence,
\beq
 C_{m_i}{}^a=\partial_a\ln y_i+\partial_a\ln v .
\eeq
Thus a compactification or runaway model gives
\beq
\lambda_{\rm string}
\longmapsto
\{C_e{}^a,C_\Lambda{}^a,C_{\hat m}{}^a,C_{m_e}{}^a,\dot\Phi^a,m_\Phi,\alpha_E{}^a\},
\label{eq:stringmap}
\eeq
where $\lambda_{\rm string}$ denotes model parameters and stabilization data. The Damour--Donoghue coefficients are then derived quantities fixed by the model. In the two-charge module,
\beq
 D_e(\lambda_{\rm string})=\alpha_E(\lambda_{\rm string})d_e(\lambda_{\rm string}),
\eeq
and
\beq
 D_{\hat m}(\lambda_{\rm string})=\alpha_E(\lambda_{\rm string})
 [d_{\hat m}(\lambda_{\rm string})-d_g(\lambda_{\rm string})].
\eeq
Define the model-predicted observable vector
\beq
 \bm\theta_s(\lambda)=
 (D_{\hat m},D_e,\zeta)_\lambda .
\eeq
The string-sector statistic is therefore
\beq
 \chi^2_{\rm string}(\lambda)=
 \chi^2_{\rm WEP+clock}[\bm\theta_s(\lambda)] .
\label{eq:chisqstring}
\eeq
The scalar range and any screening law are included in the WEP rows. A model point is excluded at a chosen confidence level when its $\Delta\chi^2$ exceeds the relevant threshold after profiling over allowed nuisance parameters. In this restricted sense, the calculation is a laboratory test of a string-derived sector.

A simple special case is a one-amplitude prediction,
\beq
 D_{\hat m}=r_s A_s,
 \qquad
 D_e=A_s,
 \qquad
 A_\alpha=\zeta_s A_s,
\label{eq:stringoneamp}
\eeq
where the string model fixes $r_s$ and $\zeta_s$ while $A_s$ is the residual scalar amplitude. Then the data give
\begin{align}
\chi^2(A_s;r_s,\zeta_s)=&
\sum_i\left[\frac{A_s X_i^{(s)}-\eta_i}{\sigma_i}\right]^2
+\left[\frac{\zeta_s A_s-A_\alpha^{\rm obs}}{\sigma_\alpha}\right]^2,
\nonumber\\
X_i^{(s)}=&\,r_s\Delta Q_{\hat m}^{(i)}+\Delta Q_e^{(i)} .
\label{eq:stringoneampchi}
\end{align}
This equation applies to any string-derived model that fixes the slope ratio and rolling normalization.

The analysis therefore has two stages. Equations~\eqref{eq:gaugefunctionmap}--\eqref{eq:chisqstring} provide the effective-field-theory interface: a compactification is mapped to low-energy slopes, a trajectory, a range, and a source charge. A genuine test of a specified string construction begins only after these quantities are computed from that construction, including its stabilization and matter sector. No explicit compactification is fitted in this paper.

The Damour--Piazza--Veneziano runaway dilaton gives a benchmark. In that model the dilaton is drawn cosmologically toward strong bare string coupling. The remaining matter couplings are tied to the inflationary density-fluctuation amplitude \cite{DamourRunaway2002}. Their estimate for composition-dependent free fall is
\beq
 \frac{\Delta a}{a}\simeq
 (5.2\times10^{-5})\alpha_{\rm had}^2,
\label{eq:dpveta}
\eeq
and for the simplest $n=2$ chaotic-inflation case this becomes approximately
\beq
 \frac{\Delta a}{a}\simeq
 1.3\left[\frac{b_F}{b_\lambda c}\right]^2\times10^{-12}.
\label{eq:dpvn2}
\eeq
For $n=4$ the corresponding estimate is approximately
\beq
 \frac{\Delta a}{a}\simeq
 0.98\left[\frac{b_F}{b_\lambda c}\right]^2\times10^{-9}.
\label{eq:dpvn4}
\eeq
Equations~\eqref{eq:dpvn2}--\eqref{eq:dpvn4} give the scale set by the original runaway-dilaton estimate. The full charge-vector likelihood then refines the comparison with the chosen compositions, signs, and nuisance parameters. The MICROSCOPE Ti--Pt row has uncertainty $2.7\times10^{-15}$. An order-unity $n=2$ runaway-dilaton coefficient is therefore far above that single-row sensitivity scale. A conservative two-sigma single-row estimate gives
\beq
 \left[\frac{b_F}{b_\lambda c}\right]^2
 \lesssim 4.2\times10^{-3}
 \qquad (n=2),
\eeq
and
\beq
 \left[\frac{b_F}{b_\lambda c}\right]^2
 \lesssim 5.5\times10^{-6}
 \qquad (n=4),
\eeq
before material-specific recalibration, signs, correlations, and scalar-sector nuisance parameters are included. Once the model slopes are specified, Eq.~\eqref{eq:chisqstring} replaces this estimate.

This step connects the fit to string-derived physics. WEP and clock data constrain any compactification or runaway model that predicts $f_A(\Phi)$, the matter slopes, the rolling trajectory, and the scalar range.

\section{Damour--Donoghue composition dictionary}

For one effective light scalar, the coupling of an atomic body $A$ is
\beq
\alpha_A\equiv\frac{\partial\ln m_A}{\partial\varphi}.
\eeq
Damour and Donoghue write this coupling as \cite{DamourDonoghue2010,DamourDonoghuePhen2010}
\begin{align}
\alpha_A=&d_g+(d_{\hat m}-d_g)Q_{\hat m}^A
+(d_{\delta m}-d_g)Q_{\delta m}^A\nonumber\\
&+(d_{m_e}-d_g)Q_{m_e}^A+d_e Q_e^A.
\label{eq:alphaA}
\end{align}
The universal gluonic term gives a common acceleration and drops out of Eotvos ratios. For a source body $E$,
\beq
\eta_{AB}^{(E)}\simeq \alpha_E(\alpha_A-\alpha_B).
\label{eq:eotvosfull}
\eeq
Keeping the two leading composition structures gives
\beq
\eta_{AB}=D_{\hat m}\Delta Q_{\hat m}^{AB}+D_e\Delta Q_e^{AB},
\label{eq:eta2}
\eeq
with
\beq
D_{\hat m}\equiv\alpha_E(d_{\hat m}-d_g),\qquad D_e\equiv\alpha_Ed_e.
\label{eq:Ddefs}
\eeq
The normalization is then fixed. $D_e$ is the source-weighted WEP coefficient. $d_e$ is the microscopic electromagnetic scalar coupling.

We use the standard approximate nuclear charges
\begin{align}
Q_{\hat m}^A&\simeq0.093-\frac{0.036}{A^{1/3}}\nonumber\\
&\quad -0.020\frac{(A-2Z)^2}{A^2}\nonumber\\
&\quad -1.4\times10^{-4}\frac{Z(Z-1)}{A^{4/3}},\nonumber\\
Q_{\delta m}^A&\simeq1.7\times10^{-3}\frac{A-2Z}{A},\nonumber\\
Q_{m_e}^A&\simeq5.5\times10^{-4}\frac{Z}{A},\nonumber\\
Q_e^A&\simeq
\left[-1.4+8.2\frac{Z}{A}\right.\nonumber\\
&\quad\left.+7.7\frac{Z(Z-1)}{A^{4/3}}\right]10^{-4}.
\label{eq:charges}
\end{align}
The fit uses the two-charge truncation in Eq.~\eqref{eq:eta2}. The terms $D_{\delta m}$ and $D_{m_e}$ can be added as extra columns. A four-coefficient fit needs additional material pairs or theory priors.

Thus the reported $D_{\hat m}$--$D_e$ correlation is a conditional property of the present two-charge fit, not a correlation established after marginalizing over all four Damour--Donoghue charges.

\section{Data set and WEP likelihood}

The WEP rows are shown in Table~\ref{tab:wepdata}. The Ti--Pt row uses the MICROSCOPE final result. The Be--Ti row uses the Eot-Wash torsion-balance result. The Rb row uses isotope free fall from atom interferometry \cite{MICROSCOPE2022,Schlamminger2008,Asenbaum2020}. The MICROSCOPE uncertainty combines the statistical and systematic parts in quadrature.

\begin{table}[t]
\caption{WEP rows used in the two-parameter fit. Charge differences are computed from Eq.~\eqref{eq:charges} using representative isotopes. A precision reanalysis should use full experimental material compositions and isotope abundances.}
\label{tab:wepdata}
\begin{ruledtabular}
\begin{tabular}{lccc}
Pair & $\eta_{AB}$ & $\Delta Q_{\hat m}$ & $\Delta Q_e$\\
\hline
Ti--Pt & $(-1.5\pm2.7)10^{-15}$ & $-2.66410^{-3}$ & $-2.00310^{-3}$\\
Be--Ti & $(0.3\pm1.8)10^{-13}$ & $-7.22810^{-3}$ & $-1.55710^{-3}$\\
$^{85}$Rb--$^{87}$Rb & $(1.6\pm3.8)10^{-12}$ & $3.31510^{-5}$ & $9.20010^{-5}$
\end{tabular}
\end{ruledtabular}
\end{table}

Define
\begin{align}
\bm\theta&=\begin{pmatrix}D_{\hat m}\\D_e\end{pmatrix},\\
X_{\rm WEP}&=10^{-3}
\begin{pmatrix}
-2.664&-2.003\\
-7.228&-1.557\\
0.03315&0.09200
\end{pmatrix}.
\label{eq:xwep}
\end{align}
With
\begin{align}
\bm y_{\rm WEP}&=10^{-15}
\begin{pmatrix}-1.5\\30.0\\1600\end{pmatrix},&
\bm\sigma_{\rm WEP}&=10^{-15}
\begin{pmatrix}2.7\\180\\3800\end{pmatrix},
\end{align}
we minimize
\begin{align}
\chisq_{\rm WEP}&=(X_{\rm WEP}\bm\theta-\bm y)^T
W (X_{\rm WEP}\bm\theta-\bm y),\\
W_{ij}&=\delta_{ij}/\sigma_i^2.
\label{eq:wepchi}
\end{align}
The analytic least-squares solution is
\beq
\hat{\bm\theta}=(X^TWX)^{-1}X^TW\bm y,
\qquad
\Sigma=(X^TWX)^{-1}.
\label{eq:leastsol}
\eeq
This gives Eq.~\eqref{eq:wepfitintro}. Diagonalizing Eq.~\eqref{eq:wepcovintro} gives one-sigma axes
\beq
0.810\times10^{-12},
\qquad
58.10\times10^{-12}.
\eeq
The long direction is the present material-composition degeneracy. The central values are consistent with zero. With three WEP rows and two fitted parameters, the statistically stable result is the covariance geometry.

The magnitude $\rho=-0.99958$ follows from the near-collinearity of the three adopted two-charge vectors and is robust as a geometric statement within that restricted design matrix. Its numerical value is not a materials-independent observable. Restoring $Q_{\delta m}$ and $Q_{m_e}$, or replacing representative isotopes by each experiment's full alloy composition and isotopic abundances, changes the design matrix and requires a new fit. The present three WEP rows have rank at most three in a four-charge analysis, so they alone cannot determine a fully unconstrained four-charge covariance.

\begin{figure}[t]
\centering
\includegraphics[width=0.96\columnwidth]{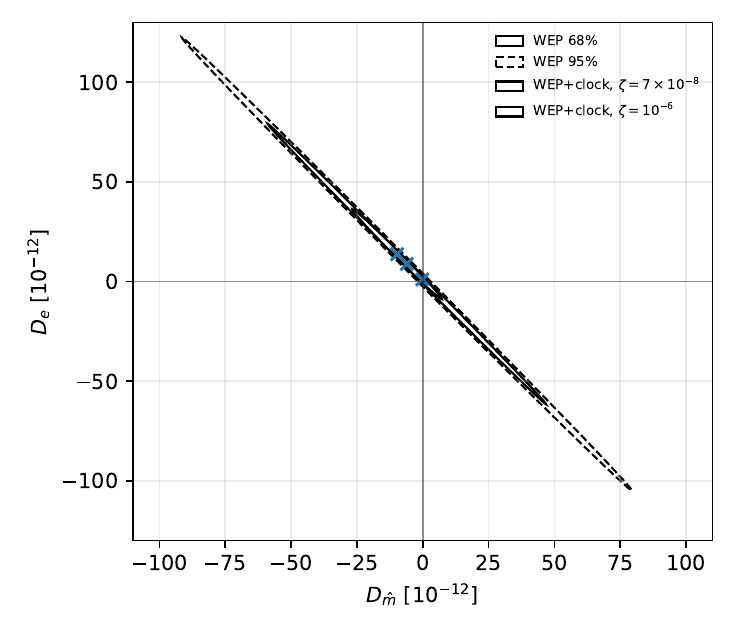}
\caption{Confidence ellipses in the $(D_{\hat m},D_e)$ plane. The WEP-only fit is elongated because the present charge vectors are nearly collinear in the two-parameter truncation. The Yb$^+$ row at $\zeta=7\times10^{-8}\,\yr^{-1}$ already reduces the ellipse, since the secular drift uncertainty is small. The $\zeta=10^{-6}\,\yr^{-1}$ case shows the regime where the clock row dominates $D_e$. Ellipses use $\Delta\chi^2=2.30$ and $5.99$ for two fitted parameters.}
\label{fig:ellipses}
\end{figure}

\section{Clock row and closure relation}

For an optical clock ratio $R_{AB}=\nu_A/\nu_B$,
\beq
\delta\ln R_{AB}=\Delta K_\alpha^{AB}\delta\ln\alpha
+\Delta K_\mu^{AB}\delta\ln\mu+\cdots .
\label{eq:clockratio}
\eeq
The older Al$^+$/Hg$^+$ comparison gave the long-baseline reference value \cite{Rosenband2008}
\beq
\left(\frac{\dot\alpha}{\alpha}\right)_{\rm Al/Hg}=(-1.6\pm2.3)\times10^{-17}\,\yr^{-1}.
\label{eq:rosenband}
\eeq
The present secular module instead uses the Yb$^+$ E3/E2 comparison \cite{Lange2021},
\beq
A_\alpha\equiv\frac{\dot\alpha}{\alpha}=(1.0\pm1.1)\times10^{-18}\,\yr^{-1}.
\label{eq:lange}
\eeq
The update shifts the equal-reach threshold by about a factor of 21.

Modern clock networks also constrain oscillatory and environmental signatures of fundamental-constant variation \cite{Barontini2022,Sherrill2023,Filzinger2023}. Those rows can be inserted in the same likelihood after the scalar time spectrum is specified. Molecular spectroscopy gives complementary constraints on mass-ratio variation \cite{Patra2020}. This paper uses the secular $\alpha$-drift row, since it gives the simplest first closure with $D_e$.

From Eq.~\eqref{eq:Ddefs},
\beq
D_e=\alpha_Ed_e.
\eeq
The same microscopic coupling gives clock drift,
\beq
A_\alpha=d_e\dot\varphi.
\eeq
So
\beq
A_\alpha=\zeta D_e,
\qquad
\zeta\equiv\frac{\dot\varphi}{\alpha_E}.
\label{eq:closure}
\eeq
Equivalently,
\beq
A_\alpha=\left(\frac{\dot\varphi}{\alpha_E}\right)(\alpha_Ed_e)=\dot\varphi d_e.
\eeq
The source charge $\alpha_E$ cancels at the microscopic level. Its role is to translate the WEP coefficient $D_e$ into the microscopic coupling $d_e$ and to define the model-dependent rolling parameter $\zeta$. The clock row constrains $D_e$ after $\zeta$ is fixed.

The reference choice $\alpha_E=10^{-3}\alpha_0$ is used only to display the illustrative conversion between $\zeta$ and $\xi/\alpha_0$. It is not a prediction of the effective theory or a direct bound on a string parameter. A model-specific value follows from the microscopic couplings, Earth composition, scalar range, and any screening prescription.

The combined statistic is
\begin{align}
\chisq(D_{\hat m},D_e;\zeta)&=
\sum_i\left[\frac{D_{\hat m}\Delta Q_{\hat m}^{(i)}+D_e\Delta Q_e^{(i)}-\eta_i}{\sigma_i}\right]^2\nonumber\\
&\quad+\left[\frac{\zeta D_e-A_\alpha^{\rm obs}}{\sigma_\alpha}\right]^2.
\label{eq:jointchi}
\end{align}
The fourth row of the design matrix is $(0,\zeta)$. The clock reaches the WEP marginal uncertainty on $D_e$ when
\beq
|\zeta|\gtrsim\frac{\sigma_\alpha}{\sigma_{D_e}^{\rm WEP}}
=\frac{1.1\times10^{-18}}{4.6459\times10^{-11}}\,\yr^{-1}
=2.37\times10^{-8}\,\yr^{-1}.
\label{eq:zetath}
\eeq
This value is an equal-reach threshold. Exclusion of $\xi/\alpha_0$ requires a predicted scalar amplitude, a prior, or a relation between $D_e$ and the other scalar slopes.

At $\zeta=7.0\times10^{-8}\,\yr^{-1}$, corresponding to $\xi/\alpha_0=1$ in Eq.~\eqref{eq:zetaintro}, the joint fit gives
\begin{align}
D_{\hat m}&=(-9.75\pm11.22)\times10^{-12},\nonumber\\
D_e&=(13.72\pm14.89)\times10^{-12},
\label{eq:naturalfit}
\end{align}
with $\chi^2_{\min}=0.190$ for two degrees of freedom. Both central values lie within one standard deviation of zero. The positive $D_e$ central value mainly reflects the mildly positive Yb$^+$ clock central value and is consistent with statistical noise. At $\zeta=10^{-6}\,\yr^{-1}$ the joint fit gives
\begin{align}
D_{\hat m}&=(-0.20\pm1.31)\times10^{-12},\nonumber\\
D_e&=(1.00\pm1.10)\times10^{-12},
\label{eq:jointstar}
\end{align}
with
\beq
\Sigma_{10^{-6}}=10^{-24}
\begin{pmatrix}
1.708&-0.908\\
-0.908&1.209
\end{pmatrix},
\qquad \rho=-0.632.
\label{eq:covstar}
\eeq
The shift from the Rosenband-based fit is large because the clock uncertainty is smaller by about a factor of 21.

\begin{figure*}[t]
\centering
\includegraphics[width=0.88\textwidth]{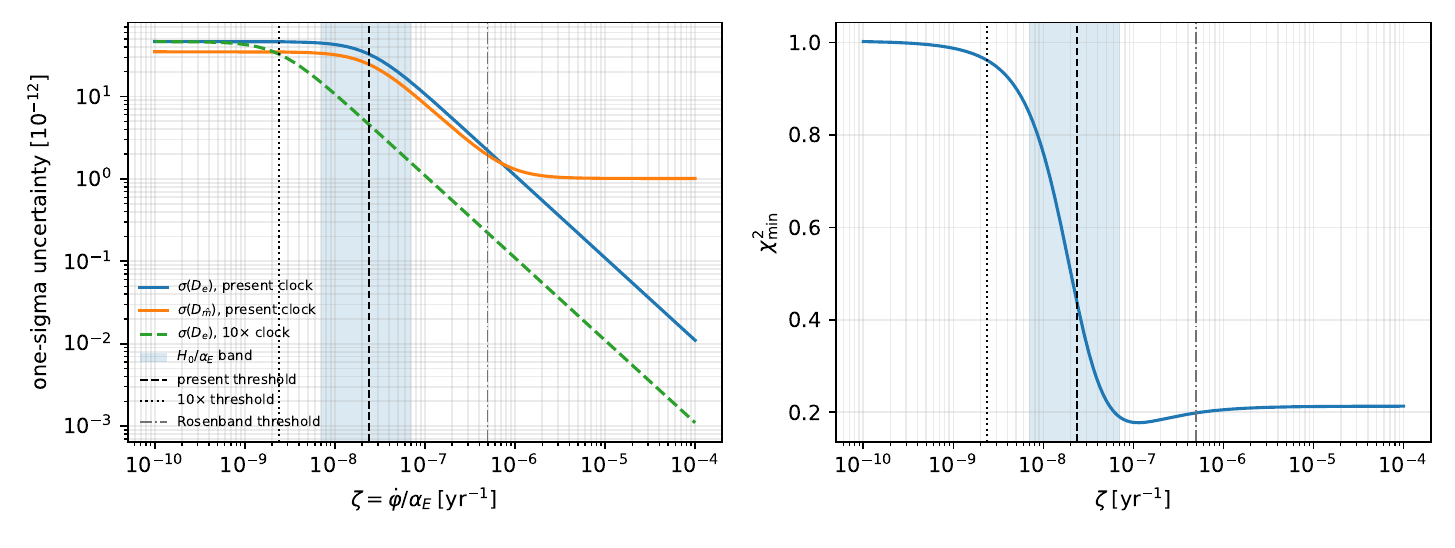}
\caption{Conditional clock--WEP sensitivity versus $\zeta=\dot\varphi/\alpha_E$. Left: one-sigma uncertainties on $D_e$ and $D_{\hat m}$ from the present Yb$^+$ secular clock row, with a null tenfold-clock forecast. The grey dash-dotted line marks the older Rosenband-based threshold. Right: minimized present-data $\chi^2$ versus $\zeta$. The shaded band corresponds to $\zeta\sim H_0/\alpha_E$ for $\alpha_E\sim10^{-3}$--$10^{-2}$ and $\xi/\alpha_0\sim1$.}
\label{fig:zeta}
\end{figure*}

\section{Rolling reference cases and one-amplitude reach}

The slow-roll reference case in Eq.~\eqref{eq:rollintro} gives Eq.~\eqref{eq:zetaintro}. Combining that equation with Eq.~\eqref{eq:zetath} gives the benchmark value at which the secular $\alpha$-clock row becomes competitive with the WEP marginal reach:
\beq
\frac{\xi}{\alpha_0}\simeq
\frac{2.37\times10^{-8}}{7.0\times10^{-8}}
\simeq0.34.
\label{eq:currentboundary}
\eeq
For a tenfold clock improvement,
\beq
\zeta_{\rm th}^{(10)}=2.37\times10^{-9}\,\yr^{-1},
\qquad
\frac{\xi}{\alpha_0}\simeq0.034.
\label{eq:futureboundary}
\eeq
The legacy Rosenband reference value would instead give
\beq
\zeta_{\rm th}^{\rm legacy}=4.95\times10^{-7}\,\yr^{-1},
\qquad
\frac{\xi}{\alpha_0}\simeq7.07.
\eeq
Table~\ref{tab:clockcompare} summarizes the shift.

\begin{table}[t]
\caption{Clock thresholds for the same WEP-only covariance. The boundary in the last column uses $\zeta=7.0\times10^{-8}(\xi/\alpha_0)\,\yr^{-1}$.}
\label{tab:clockcompare}
\begin{ruledtabular}
\begin{tabular}{lccc}
Clock input & $\sigma_\alpha$ [$\yr^{-1}$] & $\zeta_{\rm th}$ [$\yr^{-1}$] & $\xi/\alpha_0$\\
\hline
Al$^+$/Hg$^+$, 2008 & $2.3\times10^{-17}$ & $4.95\times10^{-7}$ & $7.07$\\
Yb$^+$ E3/E2, 2021 & $1.1\times10^{-18}$ & $2.37\times10^{-8}$ & $0.34$\\
Yb$^+$, 10$\times$ forecast & $1.1\times10^{-19}$ & $2.37\times10^{-9}$ & $0.034$
\end{tabular}
\end{ruledtabular}
\end{table}

A reach or exclusion statement needs one more model input. The scalar sector must provide a slope ratio, an amplitude or prior, a rolling normalization, and the scalar range or screening law. A one-amplitude case is
\beq
D_{\hat m}=rD_e,
\label{eq:rmodel}
\eeq
where $r$ is supplied by the candidate scalar sector. The statistic then becomes
\begin{align}
\chisq(D_e;r,\zeta)&=
\sum_i\left[\frac{D_e\left(r\Delta Q_{\hat m}^{(i)}+\Delta Q_e^{(i)}\right)-\eta_i}{\sigma_i}\right]^2\nonumber\\
&\quad+\left[\frac{\zeta D_e-A_\alpha^{\rm obs}}{\sigma_\alpha}\right]^2.
\label{eq:onedimchi}
\end{align}
The best-fit amplitude is analytic:
\begin{align}
D_e^{\rm bf}=&\frac{\sum_i X_i\eta_i/\sigma_i^2+\zeta A_\alpha^{\rm obs}/\sigma_\alpha^2}{\sum_i X_i^2/\sigma_i^2+\zeta^2/\sigma_\alpha^2},\nonumber\\
X_i=&r\Delta Q_{\hat m}^{(i)}+\Delta Q_e^{(i)}.
\label{eq:bestonedim}
\end{align}
The one-sigma uncertainty is
\beq
 \sigma^2(D_e)=
 \left(\sum_i\frac{X_i^2}{\sigma_i^2}
 +\frac{\zeta^2}{\sigma_\alpha^2}\right)^{-1}.
\label{eq:onedimsigma}
\eeq
A runaway dilaton, Kahler modulus, or swampland-motivated scalar enters this equation by providing $(r,\zeta)$ and an allowed range for $D_e$. The data then test the resulting point or curve. In this setting, $\xi/\alpha_0$ converts the clock-row strength into a benchmark scale.

\begin{figure}[t]
\centering
\includegraphics[width=0.98\columnwidth]{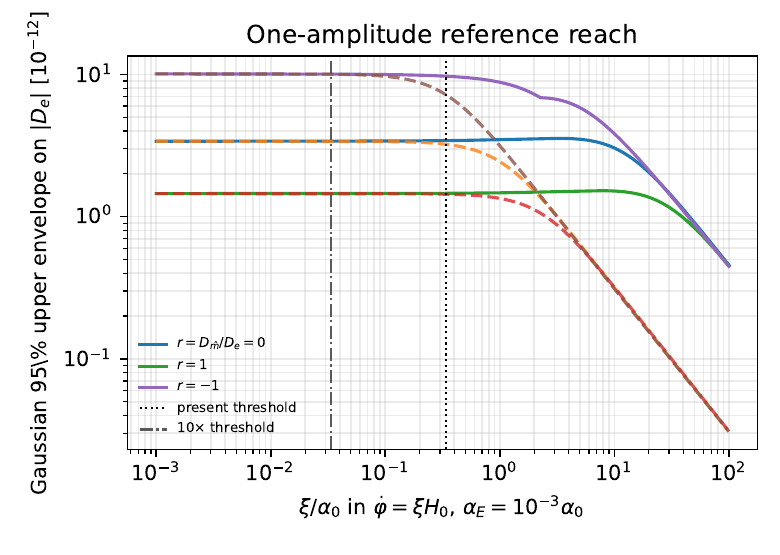}
\caption{One-amplitude reference reach. For each fixed slope ratio $r=D_{\hat m}/D_e$, the WEP+clock likelihood gives a Gaussian 95\% upper envelope on $|D_e|$. Solid curves use the present Yb$^+$ clock row. Dashed curves use a null tenfold-clock forecast. The vertical markers translate the present and tenfold equal-reach thresholds to $\xi/\alpha_0$ using $\dot\varphi=\xi H_0$ and $\alpha_E=10^{-3}\alpha_0$.}
\label{fig:reach}
\end{figure}

Figure~\ref{fig:reach} shows the resulting Gaussian 95\% upper envelope on $|D_e|$ for three representative slope ratios. For fixed $r$ and fixed $\xi/\alpha_0$, amplitudes above the curve are excluded in this one-amplitude reference case. The clock row constrains an amplitude after the slope ratio and rolling normalization are fixed.

\begin{table*}[t]
\caption{Inputs needed from several rolling-scalar model classes. The table gives model inputs used by later fits.}
\label{tab:modelclasses}
\begin{ruledtabular}
\begin{tabular}{lll}
Class & Required model input & Likelihood use\\
\hline
String/runaway dilaton & $(r,\zeta)$ and amplitude prior & One-amplitude WEP+clock test\\
Quintessence-like scalar & Rolling rate from dark-energy dynamics & Secular drift plus WEP rows\\
Oscillating scalar dark matter & Spectrum, phase, and clock $K$ coefficients & Clock-network time template\\
Generic multi-modulus sector & $\dot\varphi^a$, $\alpha_E{}^a$, and $C_X{}^a$ & Full vector likelihood\\
Higgs-sector modulus & $v(\varphi)$ and $y_i(\varphi)$ & Collider rows after slope mapping\\
Finite-range or screened scalar & $m_\varphi$, source form factor, screening profile & Range-dependent WEP source charges
\end{tabular}
\end{ruledtabular}
\end{table*}

\section{Fisher geometry and optimal composition design}

The covariance matrix also guides material choice. Let
\beq
F_0=X_{\rm WEP}^TWX_{\rm WEP},
\qquad
\Sigma_0=F_0^{-1}
\eeq
be the present Fisher matrix and covariance in the two-parameter truncation. A future Eotvos row with charge vector
\beq
\bm q=(\Delta Q_{\hat m},\Delta Q_e)
\eeq
and uncertainty $\sigma_\eta^{\rm new}$ changes the Fisher matrix by
\beq
F_{\rm new}=F_0+\frac{\bm q\bm q^T}{(\sigma_\eta^{\rm new})^2}.
\label{eq:fisherupdate}
\eeq
The long-axis shrink factor is
\beq
R_{\rm maj}(\bm q)=
\sqrt{\frac{\lambda_{\rm maj}(\Sigma_0)}
{\lambda_{\rm maj}(F_{\rm new}^{-1})}}.
\label{eq:rmaj}
\eeq
This number asks which composition contrast best breaks the present degeneracy.

We scanned representative isotopes with the charge formula in Eq.~\eqref{eq:charges}. The scan gives charge-geometry targets for later mission design. The entries are target directions in charge space. A final experimental analysis should use actual alloy composition, isotope abundances, sign conventions, charge uncertainties, and engineering constraints. In particular, a MICROSCOPE-level Be--Ti-like row means a precision-upgraded light--medium charge contrast; a finished space mission would also require material and engineering studies. For a reference new-row precision equal to the final MICROSCOPE Ti--Pt uncertainty,
\beq
\sigma_\eta^{\rm new}=2.7\times10^{-15},
\eeq
Table~\ref{tab:design} gives the leading charge-vector choices. The present long axis is set by the lower precision of the existing Be--Ti row. A high-precision light--medium contrast gives much more information in the two-coefficient plane than another Ti--Pt-like row.

The factors in Table~\ref{tab:design}, including $27.8$, are idealized Fisher forecasts conditional on one independent new row with $\sigma_\eta^{\rm new}=2.7\times10^{-15}$, the two-charge truncation, and representative isotopes. They omit engineering constraints, composition uncertainties, systematic-error correlations, and any additional scalar charges.

\begin{table}[t]
\scriptsize
\caption{Fisher-design scan for one additional WEP row with $\sigma_\eta^{\rm new}=2.7\times10^{-15}$. $R_{\rm maj}$ is the shrink factor of the present long covariance axis. Entries use representative isotope charge vectors and indicate charge-geometry targets.}
\label{tab:design}
\begin{ruledtabular}
\begin{tabular}{lcccc}
Pair & $\Delta Q_{\hat m}$ & $\Delta Q_e$ & $R_{\rm maj}$ & long axis\\
\hline
Be--Ti & $-7.228\times10^{-3}$ & $-1.557\times10^{-3}$ & $27.76$ & $2.09$\\
Be--Al & $-5.347\times10^{-3}$ & $-1.020\times10^{-3}$ & $27.51$ & $2.11$\\
Be--Cr & $-7.482\times10^{-3}$ & $-1.710\times10^{-3}$ & $27.14$ & $2.14$\\
Be--Fe & $-7.707\times10^{-3}$ & $-1.859\times10^{-3}$ & $26.51$ & $2.19$\\
Be--Rb & $-8.589\times10^{-3}$ & $-2.151\times10^{-3}$ & $26.32$ & $2.21$\\
Ti--Pt repetition & $-2.664\times10^{-3}$ & $-2.003\times10^{-3}$ & $1.00$ & $58.09$\\
Rb isotope & $3.315\times10^{-5}$ & $9.200\times10^{-5}$ & $1.53$ & $38.05$
\end{tabular}
\end{ruledtabular}
\end{table}

\begin{figure*}[t]
\centering
\includegraphics[width=0.92\textwidth]{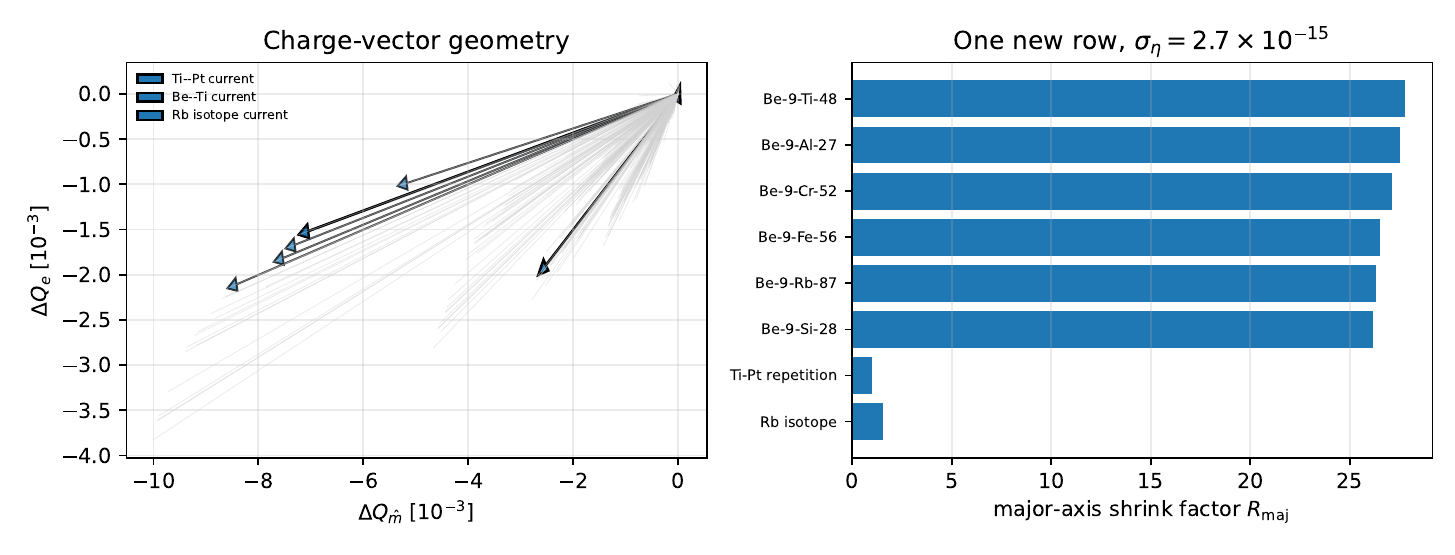}
\caption{Composition geometry in the two-charge Damour--Donoghue plane. Left: current and candidate charge vectors. The Ti--Pt vector is already measured with high precision. The Be--Ti direction has lower precision and controls the long covariance axis. Right: major-axis shrink factor for one hypothetical MICROSCOPE-level future row. A high-precision Be--light/medium contrast breaks the current degeneracy. A second Ti--Pt-like row mainly improves the already tight direction.}
\label{fig:design}
\end{figure*}

The lesson is geometric. A new measurement along an already precise charge direction adds little to the long axis. In this data set, Be--Ti-like charge vectors are the efficient direction.

\section{Scope and limitations}

The calculation makes the scalar connection testable at low energy. WEP tests constrain $\alpha_Ed_e$. Clocks constrain $d_e\dot\varphi$. The ratio $\dot\varphi/\alpha_E$ must come from the candidate scalar sector.

The limits of the calculation are clear. First, the closure relation is a one-direction statement. Generic multi-field models require the full vectors $\dot\varphi^a$, $\alpha_E{}^a$, and $C_e{}^a$. Second, the WEP rows assume a scalar that is effectively long-ranged over the experimental source scale. Finite-range or screened fields require range-dependent source charges. Third, $\alpha_E=10^{-3}\alpha_0$ is a benchmark conversion. In a complete model, $\alpha_E$ is computed from Earth-like matter and from the same microscopic couplings that enter the WEP coefficients.

The material charges use representative isotopes. A final experimental analysis should use the actual MICROSCOPE Ti alloy and Pt/Rh alloy, natural isotope abundances where relevant, sign conventions for each reported $\eta_{AB}$, and charge uncertainties. The two-charge truncation is a practical leading approximation. Adding $Q_{\delta m}$ and $Q_{m_e}$ requires additional material pairs or theory priors.

Consequently, the manuscript reports conditional constraints on rolling-scalar parameter combinations; it does not claim universal constraints on all string-derived dilaton or modulus sectors. The strong two-charge anti-correlation is an informative diagnostic for the present reduced analysis, while a precision bound on a specified microscopic model should be obtained from a full-composition, full-charge likelihood.

The word ``test'' has this restricted meaning. Present clocks make the $\zeta$ row comparable to WEP within the illustrative slow-roll range. They exclude a rolling-dilaton model when that model predicts an amplitude and a slope ratio. The connection to string-derived physics is then direct: a compactification, dilaton runaway, or swampland-motivated construction can be compared with the likelihood after it predicts the low-energy scalar data.

\section{Conclusion}

Rolling scalar sectors should be tested with a common slope likelihood. Clock, WEP, molecular, and composition limits measure different projections of the same microscopic data once a model fixes the scalar trajectory. The calculation builds that likelihood, runs the long-range WEP plus secular $\alpha$-clock module, and uses the covariance to identify future composition directions.

The new result is a reproducible conditional likelihood that places current WEP data and the updated Yb$^+$ secular $\alpha$-drift result in the same scalar-slope parameter space. Relative to earlier separate varying-constant and equivalence-principle analyses, the manuscript (i) writes the field-space condition required for their common closure relation, (ii) quantifies the conditional clock--WEP equal-reach diagnostic using the current Yb$^+$ input, and (iii) uses the resulting covariance to identify composition directions that would break the present two-charge degeneracy. These are low-energy phenomenological results; they become tests of a particular string-derived model only after that model supplies its compactification-dependent slopes, stabilization data, source charge, scalar range, and trajectory.

The WEP-only Damour--Donoghue fit gives
\begin{align}
D_{\hat m}&=(-6.06\pm34.91)\times10^{-12},\\
D_e&=(8.80\pm46.46)\times10^{-12},
\end{align}
with $\rho=-0.99958$. The large anti-correlation is the present WEP degeneracy. Adding the Yb$^+$ clock row through $A_\alpha=\zeta D_e$ gives a conditional family of fits. At $\zeta=7\times10^{-8}\,\yr^{-1}$, corresponding to $\xi/\alpha_0=1$ for $\dot\varphi=\xi H_0$ and the benchmark $\alpha_E=10^{-3}\alpha_0$, the joint result is
\begin{align}
D_{\hat m}&=(-9.75\pm11.22)\times10^{-12},\\
D_e&=(13.72\pm14.89)\times10^{-12}.
\end{align}
The mild positive central value is consistent with statistical noise and mainly follows the positive central value of the Yb$^+$ drift input.

For the stated benchmark conversion, the present equal-information boundary is $\xi/\alpha_0\simeq0.34$. A tenfold clock improvement moves it to $0.034$. These numbers mark when the clock row competes with WEP in the $D_e$ direction. They mark equal information. Model exclusion follows from the one-amplitude likelihood after a model supplies $r=D_{\hat m}/D_e$, the amplitude range, the source coupling, and the scalar trajectory.

The Fisher scan gives the experimental lesson. In the representative-isotope two-charge model, a high-precision Be--Ti-like contrast reduces the long WEP axis by a factor $27.8$. Repeating the Ti--Pt direction leaves that axis almost unchanged. The next gain comes from a charge direction that breaks the present degeneracy together with improved precision.

The result is a reproducible low-energy likelihood for rolling scalar sectors. A correlated deviation across clocks and WEP experiments would point to vacuum-dependent Standard-Model parameters. Continued null results reduce a well-defined scalar-slope space. A particular ultraviolet theory supplies its slope ratios, source coupling, scalar range, and scalar trajectory before comparison with the likelihood.

\section*{Acknowledgments}

The author sincerely thanks the anonymous referees for their careful and constructive reports. Their comments led to a sharper separation between the effective low-energy likelihood, its explicit one-direction and long-range assumptions, and a model-specific test of a string-derived scalar sector.

\appendix

\section{Numerical reconstruction}

Table~\ref{tab:fits} gives the numerical results used in the figures. All $D$ values and uncertainties are in units of $10^{-12}$.

\begin{table}[h]
\caption{Best fits for selected values of $\zeta$ using the Yb$^+$ clock row.}
\label{tab:fits}
\begin{ruledtabular}
\begin{tabular}{lccccc}
Input & $D_{\hat m}$ & $\sigma_{D_{\hat m}}$ & $D_e$ & $\sigma_{D_e}$ & $\chisq_{\min}$\\
\hline
WEP only & $-6.06$ & $34.91$ & $8.80$ & $46.46$ & $0.177$\\
$\zeta=2.37\times10^{-8}$ & $-18.61$ & $24.69$ & $25.52$ & $32.85$ & $0.436$\\
$\zeta=7.0\times10^{-8}$ & $-9.75$ & $11.22$ & $13.72$ & $14.89$ & $0.190$\\
$\zeta=10^{-6}$ & $-0.20$ & $1.31$ & $1.00$ & $1.10$ & $0.205$\\
$\zeta=10^{-4}$ & $0.55$ & $1.01$ & $0.010$ & $0.011$ & $0.213$
\end{tabular}
\end{ruledtabular}
\end{table}

The confidence ellipses in Fig.~\ref{fig:ellipses} use
\beq
(\bm\theta-\hat{\bm\theta})^T\Sigma^{-1}(\bm\theta-\hat{\bm\theta})=\Delta\chisq,
\eeq
with $\Delta\chisq=2.30$ and $5.99$ for 68\% and 95\% confidence in two parameters.

\section{Four-charge rank check}

The executed WEP fit uses the two leading Damour--Donoghue charge columns. If the two subleading columns $(Q_{\delta m},Q_{m_e})$ are restored, the present three WEP rows form a $3\times4$ design matrix. Using the same representative isotopes, the correct unweighted four-column rows are
\beq
\begin{pmatrix}
-2.6638&-2.0026&-0.1983& 0.0321\\
-7.2282&-1.5571& 0.0472&-0.0076\\
 0.0331& 0.0920&-0.0340& 0.0055
\end{pmatrix}10^{-3},
\eeq
where the columns are $(\Delta Q_{\hat m},\Delta Q_e,\Delta Q_{\delta m},\Delta Q_{m_e})$. The matrix rank is at most three. This confirms that a fully free four-charge fit needs more material contrasts or theory priors.


\begin{thebibliography}{99}

\bibitem{Vafa2005}
C. Vafa, The string landscape and the swampland, \arxivlink{hep-th/0509212}.

\bibitem{OoguriVafa2007}
H. Ooguri and C. Vafa, On the geometry of the string landscape and the swampland, Nucl. Phys. B \textbf{766}, 21 (2007), \doilink{10.1016/j.nuclphysb.2006.10.033}.

\bibitem{Palti2019}
E. Palti, The swampland: introduction and review, Fortschr. Phys. \textbf{67}, 1900037 (2019), \doilink{10.1002/prop.201900037}.

\bibitem{DamourRunaway2002}
T. Damour, F. Piazza, and G. Veneziano, Runaway dilaton and equivalence principle violations, Phys. Rev. Lett. \textbf{89}, 081601 (2002), \doilink{10.1103/PhysRevLett.89.081601}.


\bibitem{KaplanWise2000}
D. B. Kaplan and M. B. Wise, Couplings of a light dilaton and violations of the equivalence principle, JHEP \textbf{08}, 037 (2000), \doilink{10.1088/1126-6708/2000/08/037}.

\bibitem{GKP2002}
S. B. Giddings, S. Kachru, and J. Polchinski, Hierarchies from fluxes in string compactifications, Phys. Rev. D \textbf{66}, 106006 (2002), \doilink{10.1103/PhysRevD.66.106006}.

\bibitem{KKLT2003}
S. Kachru, R. Kallosh, A. Linde, and S. P. Trivedi, de Sitter vacua in string theory, Phys. Rev. D \textbf{68}, 046005 (2003), \doilink{10.1103/PhysRevD.68.046005}.

\bibitem{Burgess2020}
C. P. Burgess, \textit{Introduction to Effective Field Theory}, Cambridge University Press (2020), \doilink{10.1017/9781139048040}.

\bibitem{Uzan2011}
J.-P. Uzan, Varying constants, gravitation and cosmology, Living Rev. Relativity \textbf{14}, 2 (2011), \doilink{10.12942/lrr-2011-2}.

\bibitem{Uzan2025}
J.-P. Uzan, Fundamental constants: from measurement to the universe, a window on gravitation and cosmology, Living Rev. Relativity \textbf{28}, 6 (2025), \doilink{10.1007/s41114-025-00059-y}.

\bibitem{ATLAS2022}
G. Aad \textit{et al.} (ATLAS Collaboration), A detailed map of Higgs boson interactions by the ATLAS experiment ten years after the discovery, Nature \textbf{607}, 52 (2022), \doilink{10.1038/s41586-022-04893-w}.

\bibitem{CMS2022}
A. Tumasyan \textit{et al.} (CMS Collaboration), A portrait of the Higgs boson by the CMS experiment ten years after the discovery, Nature \textbf{607}, 60 (2022), \doilink{10.1038/s41586-022-04892-x}.

\bibitem{DamourDonoghue2010}
T. Damour and J. F. Donoghue, Equivalence principle violations and couplings of a light dilaton, Phys. Rev. D \textbf{82}, 084033 (2010), \doilink{10.1103/PhysRevD.82.084033}.

\bibitem{DamourDonoghuePhen2010}
T. Damour and J. F. Donoghue, Phenomenology of the equivalence principle with light scalars, Class. Quantum Grav. \textbf{27}, 202001 (2010), \doilink{10.1088/0264-9381/27/20/202001}.

\bibitem{MICROSCOPE2022}
P. Touboul \textit{et al.}, MICROSCOPE mission: final results of the test of the equivalence principle, Phys. Rev. Lett. \textbf{129}, 121102 (2022), \doilink{10.1103/PhysRevLett.129.121102}.

\bibitem{Schlamminger2008}
S. Schlamminger, K.-Y. Choi, T. A. Wagner, J. H. Gundlach, and E. G. Adelberger, Test of the equivalence principle using a rotating torsion balance, Phys. Rev. Lett. \textbf{100}, 041101 (2008), \doilink{10.1103/PhysRevLett.100.041101}.

\bibitem{Asenbaum2020}
P. Asenbaum, C. Overstreet, M. Kim, J. Curti, and M. A. Kasevich, Atom-interferometric test of the equivalence principle at the $10^{-12}$ level, Phys. Rev. Lett. \textbf{125}, 191101 (2020), \doilink{10.1103/PhysRevLett.125.191101}.

\bibitem{Rosenband2008}
T. Rosenband \textit{et al.}, Frequency ratio of Al$^+$ and Hg$^+$ single-ion optical clocks; metrology at the 17th decimal place, Science \textbf{319}, 1808 (2008), \doilink{10.1126/science.1154622}.

\bibitem{Lange2021}
R. Lange \textit{et al.}, Improved limits for violations of local position invariance from atomic clock comparisons, Phys. Rev. Lett. \textbf{126}, 011102 (2021), \doilink{10.1103/PhysRevLett.126.011102}.

\bibitem{Barontini2022}
G. Barontini \textit{et al.}, Measuring the stability of fundamental constants with a network of clocks, EPJ Quantum Technol. \textbf{9}, 12 (2022), \doilink{10.1140/epjqt/s40507-022-00130-5}.

\bibitem{Sherrill2023}
N. Sherrill \textit{et al.}, Analysis of atomic-clock data to constrain variations of fundamental constants, New J. Phys. \textbf{25}, 093012 (2023), \doilink{10.1088/1367-2630/aceff6}.

\bibitem{Filzinger2023}
M. Filzinger \textit{et al.}, Improved limits on the coupling of ultralight bosonic dark matter to photons from optical atomic clock comparisons, Phys. Rev. Lett. \textbf{130}, 253001 (2023), \doilink{10.1103/PhysRevLett.130.253001}.

\bibitem{Patra2020}
S. Patra \textit{et al.}, Proton-electron mass ratio from laser spectroscopy of HD$^+$ at the part-per-trillion level, Science \textbf{369}, 1238 (2020), \doilink{10.1126/science.aba0453}.

\end{thebibliography}
\end{document}